%% file: 0_main.tex
\documentclass[runningheads]{llncs}

\usepackage{graphicx}
\usepackage[utf8]{inputenc}
\usepackage[T1]{fontenc}
\usepackage[numbers]{natbib}
\usepackage{xcolor}
\usepackage{tikz}

\begin{document}

\title{End-to-end verifiable voting for developing countries - what's hard in Lausanne is harder still in Lahore}

%\thanks{This work has been produced under ..}}
%
\titlerunning{End-to-end verifiable voting for developing countries}
\author{Hina Binte Haq\inst{1}$^,$\inst{2} \and
Syed Taha Ali \inst{2} \and
Ronan McDermott\inst{3}}
\authorrunning{Hina Binte Haq et al.}
\institute{National University of Computer and Emerging Sciences, Islamabad, Pakistan. \email{hina.haq@nu.edu.pk} \and
School of Electrical Engineering and Computer Sciences (SEECS), National
University of Sciences and Technology (NUST), Islamabad, Pakistan.
\email{taha.ali@seecs.edu.pk}\\ \and
MCDIS \email{ronan@mcdis}}
\maketitle              % typeset the header of the contribution
\begin{abstract}
%In recent years end-to-end verifiable voting (E2EVV) has emerged as a promising new paradigm to conduct evidence based elections. However, E2EVV systems thus far have primarily been designed for the developed world and the underlying assumptions in the design of these systems do not translate to the developing world, and may even act as potential barriers to adoption of E2EVV systems. This is unfortunate, as developing countries account for 80\% of the world's population, and given their economic and socio-political dilemmas and their long histories of contentious elections, they stand to benefit most from this exciting new paradigm. In this paper, we propose alternate principles and strategies more relevant to the developing world and we discuss the myriad resulting concerns, the social, political, technical, operational, and human challenges that researchers and policymakers need to address. Our goal is to outline a broader research agenda for the community to successfully adapt E2EVV voting systems to developing countries.

% new abstract for practitioner's track [TA]

In recent years end-to-end verifiable voting (E2EVV) has emerged as a promising new paradigm to conduct evidence-based elections. However, E2EVV systems thus far have primarily been designed for the developed world and the fundamental assumptions underlying the design of these systems do not readily translate to the developing world, and may even act as potential barriers to adoption of these systems. This is unfortunate because developing countries account for 80\% of the global population, and given their economic and socio-political dilemmas and their track record of contentious elections, these countries arguably stand to benefit most from this exciting new paradigm. In this paper, we highlight various limitations and challenges in adapting E2EVV systems to these environments, broadly classed across social, political, technical, operational, and human dimensions. We articulate corresponding research questions and identify significant literature gaps in these categories. We also suggest relevant strategies to aid researchers, practitioners, and policymakers in visualizing and exploring solutions that align with the context and unique ground realities in these environments. Our goal is to outline a broader research agenda for the community to successfully adapt E2EVV voting systems to developing countries.

\end{abstract}

%\begin{multicols}{2}

\input{1_introduction}
\input{2_background}
\input{3_motivation}
\input{4_challenges}
\input{5_wayforward}

\small
\section*{Acknowledgement}
This research was supported by the ‘Research for Social Transformation \& Advancement’ (RASTA), a Pakistan Institute of Development Economics (PIDE) initiative, through Competitive Grants Programme Award [Grant No. CGP-01-127/2021].

\small
\bibliographystyle{splncs04}
\bibliography{Bibliography}

%\end{multicols}
%
\end{document}

%% file: 1_introduction.tex
\section{Introduction}
\label{sec:introduction}
% brief intro of E2EvV
In recent years end-to-end verifiable voting (E2EVV) has emerged as a revolutionary new paradigm to enable secure and transparent elections \cite{ali2016overview}. E2EVV voting systems preclude implicit trust in administrators, polling staff, and voting machines, and instead make voters themselves active participants in auditing the election and certifying its results - ``the Holy Grail for electronic voting'' \cite{Ummelas2017Jul}. 
% endorsements + pilots + deployments + commercialization 
These systems are backed by expert bodies \cite{national2018securing} and have been piloted in numerous small-scale mock elections and pilots \cite{hao2020end} \cite{zagorski2013remotegrity} \cite{sherman2010scantegrity}, some large-scale politically binding elections - Australia in 2016 \cite{burton2016vvote} and, most notably, nationwide deployment in the Estonian parliamentary elections in 2019 \cite{estoniaverifying}. This technology is also on the cusp of commercialization\cite{microsoft}.

% intro to our angle + assumptions
As these systems transition to the mainstream, we consider it an opportune moment to revisit gaps in the research literature, particularly with regards to deploying these systems in developing countries - environments which would arguably benefit most from the superior integrity and trust guarantees offered by these systems. E2EVV systems thus far have primarily been designed for the developed world, where it is largely assumed that there is a sufficient infrastructure for elections, voters are largely literate and relatively technically sophisticated, and dispute resolution mechanisms are reliable and effective. These assumptions do not necessarily translate to the developing world and may even act as potential barriers to adoption of E2EVV systems in these countries. 

% summary of motivation section
To motivate this study, we consider the fact that the overwhelming majority of the global population - approximately 80\% - hails from the developing world \cite{popbycountry}. Distrust in democracy and electoral processes runs high in many of these countries \cite{norris2015contentious}, and election fraud has resulted in mass protests \cite{Lopez2017}, political deadlock \cite{shah2016democracy}, and violence \cite{Goodman2021Feb}. Some of these countries have introduced electronic voting systems, and, in several, results have proved controversial \cite{cheeseman2018digital}. %refs replaced 

% need for E2EVV voting + calls %contributions
We believe that E2EVV systems, with their potential to restore trust and confidence in electoral processes, have a vital role to play in the developing world. In recent years there have even been public calls to explore application of this technology in countries including Brazil \cite{aranha2018good}, Pakistan \cite{haq2019pakistan} and India \cite{cce_report}. \\ In this paper, we make the following contributions:
%\noindent \tikz \fill[black] (0, 0) circle [radius=0.075];  We contend that certain fundamental assumptions implicit in the design of E2EVV systems developed thus far conflict with ground realities in developing countries. To make these assumptions explicit, we highlight the manifold limitations and challenges in adapting E2EVV systems to these new environments. \\
  %  \noindent \tikz \fill[black] (0, 0) circle [radius=0.075];  This reorientation opens up significant new ground. We identify and elaborate various social, political, technical, operational, and human concerns specific to E2EVV systems in developing countries and frame specific research questions. \\
   % \noindent \tikz \fill[black] (0, 0) circle [radius=0.075];  We suggest potential strategies for the way forward based on relevant trends and success stories in developing countries as well as re-purposing solutions from research literature. Our goal is to aid the community to devise solutions that are appropriate to the unique ground realities of the developing world. 
    \begin{itemize}
    \item We contend that certain fundamental assumptions implicit in the design of E2EVV systems developed thus far conflict with ground realities in developing countries. To make these assumptions explicit, we highlight the manifold challenges in adapting E2EVV systems to these environments.
    \item This reorientation opens up significant new ground. We identify various social, political, technical, operational, and human concerns specific to E2EVV systems in developing countries and frame specific research questions.
    \item We suggest potential strategies for the way forward based on relevant trends and success stories in developing countries as well as re-purposing solutions from research literature. Our goal is to aid the community to devise solutions that are appropriate to the unique ground realities of the developing world. 

    \end{itemize}

% how our work relates to related work
There is a considerable body of research on the challenges of deploying election technology in the developing world \cite{Hapsara2017Jan} \cite{agbesi2018adoption}. To the best of our knowledge, we are the first to focus specifically on adapting \textit{E2EVV systems} to the socio-political realities and infrastructure in these countries. One of our primary contributions is a detailed review of the literature, election experiences, and news reports from the developing world.

We believe this is a critical research gap, addressing which requires close collaboration between researchers, technologists, practitioners, and policymakers. We hope our paper stimulates exciting and impactful new thinking and research and extends the benefits of E2EVV voting systems to the developing world.

%% file: 2_background.tex
\section{Background and Prior Work}
\label{sec:background}

In this section, we briefly describe E2EVV voting systems, we motivate the case for their application in the developing world, and we discuss prior work.

\subsection{End-to-End Verifiable Voting Systems}

% very very high level desc.
E2EVV systems are a promising new class of voting systems which offer voters the benefits of automation, ease of vote-casting and quick reporting of results, along with stringent cryptographic guarantees of voter privacy and correct computation of the tally. Numerous such systems have been proposed over the years for precinct-based and Internet voting \cite{ali2016overview}. We summarize next the high-level workings of a representative system to convey to the reader a non-technical and intuitive understanding of end-to-end verifiable voting.

% vote casting
On the day of elections, our citizen, say Alice, arrives at a polling station and identifies herself as an eligible voter. She makes her candidate choice on a voting terminal. The machine records and encrypts her vote and issues her a \textbf{printed receipt}, bearing a unique serial number and a cryptographic commitment to her vote. This receipt allows her to later verify that her vote has been correctly processed and counted. However, the receipt does not reveal Alice's choice of candidate and she cannot use it to sell her vote.

% cast as intended
However, Alice may suspect the machine is malfunctioning or has been tampered with. In this case, she avails an option to force the machine to reveal the cryptographic parameters it used to encrypt her vote. This step effectively `spoils' her ballot but allows her to double-check that the machine is operating correctly. She can repeat this step several times until she is ready to cast her vote. In the parlance of E2EVV systems, Alice is now confident that \textbf{her vote has been cast as intended}.

% recorded as cast + large scale fraud detection
When polls close, election staff post copies of all receipts online. Alice uses the serial number to navigate to her receipt. If anyone has tampered with her vote, she can detect it by comparing the receipt to the physical copy she holds in her hand, and can file a complaint using her physical receipt as hard evidence. This gives her confidence that \textbf{her vote has been recorded as cast}. %Statistical analyses indicate that if a mere fraction of voters were to check their votes in this way, they would detect any attempt at large scale vote tampering with very high likelihood.

% counted as recorded
E2EVV systems usually employ two key techniques to tally results in a privacy-preserving manner: systems such as Pr\^et \'a Voter \cite{ryan2009pret} and Scantegrity \cite{chaum2008scantegrity} rely on mixnets to anonymize and decrypt cast votes which are then added. The other approach, exemplified by STAR-Vote \cite{bell2013star}, employs homomorphic encryption to aggregate encrypted votes and decrypt only the tally. Both processes offer voters and observers cryptographic proofs of correct operation. Alice can use these proofs to verify that \textbf{her vote has been tallied as recorded}.

% summing it up
These three guarantees span all critical steps of the election life-cycle, and empower users to verify the integrity of the process for themselves. By empowering voters to verify the integrity of the process themselves, E2EVV represents a dramatic improvement over traditional `black box' voting machines. 

Because of space constraints, we have eschewed technical details and refer the reader to \cite{ali2016overview} for the same.
% the paper trail - remove if space is less, not vitally necessary?
%Some E2EVV voting systems also print a physical copy of every cast vote, which voters verify and deposit in a ballot box [Hina, ref - starvote, Wombat?]. This constitutes a paper trail that can be independently audited to verify that the system is working correctly.

\subsection{Prior Work}

There is little work specifically on E2EVV systems and the developing world. Exceptions include E2EVV systems Threeballot, Twin \cite{santin2008three} and Aperio \cite{essex2010aperio}, which are geared to provide verifiability in ``minimally equipped election environments''. These systems are of considerable interest due to their novel paper-based design which precludes any use of technology or cryptography. Unfortunately there has been no effort to adapt or pilot these systems in a developing country. Moreover, they lack the highly desirable benefits of automation such as less spoiled votes, prompt reporting of results, and enfranchisement of marginal communities \cite{Somanathan2019May}.

% electronic voting
There is, however, a considerable body of research on the application and challenges of electronic voting in the developing world which is relevant to our purposes. This includes feasibility studies \cite{maphunye2019feasibility}, holistic frameworks \cite{osho2016framework}, cost benefit analyses \cite{okorocost}, and adoption studies \cite{alomari2016voting} \cite{agbesi2018adoption}. Some studies focus on specific topics, such as economic determinants of voter behaviour \cite{oganesyan2014economic} or technical concerns \cite{akinyokun2020secure}. There are case studies on e-voting in individual countries \cite{aranha2018good} \cite{inuwa2015impact}, and efforts to adapt insightful metrics, such as the E-Voting Readiness Index \cite{maletic2019scaffolding}. %There is also a wealth of supplementary material for practitioners \cite{reynolds2008electoral} \cite{goldsmith2013implementing}. \cite{aljarrah2016voting}

This literature contains several findings, which generalize to E2EVV systems and which, as we noted earlier, clash directly with the ground realities in technologically advanced countries. For instance, in the developing world resources and infrastructure for elections is commonly inadequate \cite{maphunye2019feasibility} \cite{osho2016framework} and countries often face severe financial constraints \cite{aranha2018good} \cite{maphunye2019feasibility} \cite{okorocost}. Governments may also lack technical, administrative, and operational capacity to conduct elections \cite{maphunye2019feasibility} \cite{aranha2018good}. Election management bodies often face issues of autonomy \cite{salimonu2013adoption} \cite{ahmad2015issues}. The political environment may be volatile which affects conduct of elections \cite{okorocost}. Corruption and election fraud are systemic and transparency and accountability are lacking \cite{oganesyan2014economic} \cite{okorocost} \cite{aranha2018good}. Voters may be suspicious of election technology and acceptance and adoption can be problematic \cite{alomari2016voting} \cite{agbesi2018adoption} \cite{chauhan2018acceptance}. 

Any effort to introduce E2EVV systems has to engage with these fundamental ground realities.  We explore these themes in more detail in the following sections.

% down to 2, more recent
%1. To establish theoretical common ground and adequate strategic design \cite{Hapsara2017Jan}
%2. Lack of Infrastructure \cite{ahmad2015issues} \cite{maphunye2019feasibility} \cite{osho2016framework}
%3. Systemic corruption, lack of transparency and accountability \cite{ahmad2015issues} \cite{oganesyan2014economic} \cite{okorocost} \cite{herstatt2014india} \cite{aranha2018good}
%4. Trust, acceptance and adoption  \cite{ahmad2015issues} \cite{omotayo2021adoption} \cite{alomari2016voting} \cite{agbesi2018adoption} \cite{chauhan2018acceptance} \cite{avgerou2013explaining}
%5. Lack of technical, operational and administrative capacity. \cite{salimonu2013adoption} \cite{maphunye2019feasibility} \cite{herstatt2014india} \cite{aranha2018good}
%6. Lack of EMB autonomy \cite{salimonu2013adoption} \cite{ahmad2015issues}
%7. Volatile Political Environment \cite{okorocost}
%8. Severe Financial Constraints \cite{aranha2018good}  \cite{maphunye2019feasibility} \cite{okorocost}

%% file: 3_motivation.tex
\section{Motivation}

% commencing the population argument
The vast majority of the world's population, a staggering 80\%, live in the developing world \cite{popbycountry}, covering the landmass of Africa, Latin America, and much of the Middle East and Asia. These include six of the world's ten most populous countries, namely India, Indonesia, Pakistan, Brazil, Nigeria and Bangladesh \cite{popbycountry}. Despite significant variance in size, demography, history, and culture, many of these countries face similar economic and socio-political problems: widespread poverty and inequality, low literacy, poor governance, systemic corruption, and dependence on foreign institutions \cite{west2002diverse}. Elections are routinely contentious and frequently result in political deadlock, street protests, and violence \cite{norris2015contentious}.

% examples - deadlock, street protests
For instance, in Pakistan, major rigging allegations in the general elections of 2013 resulted in mass protests and a public sit-in by a major opposition party \cite{sitinloss}. Likewise, in 2014, the Bangladesh saw a national strike lasting 85 days and a violent crackdown on opposition workers. Opposition parties boycotted the election and over half the seats went uncontested \cite{shah2016democracy}. Opposition boycotts, street protests, and civil unrest, also featured recently in Venezuela \cite{JoeSterling2017july}.

% extreme examples - violence, overthrow, coups
Perceptions of rigging can also trigger major political disruptions and take on life-threatening consequences. Allegations of ``terrible fraud'' were a key justification for the recent military coup in Myanmar, where army officials claim to have identified some 10.5 million irregularities in the voter list used in the last general elections \cite{Goodman2021Feb}. The violence following the coup resulted in over 700 deaths and more than 3,300 people detained.

% move to technology, % the positives
Technology has often been introduced, with mixed results, to resolve this issue of trust. Reported improvements in India include a significant decline in electoral fraud, a more competitive electoral process, and increased participation of marginalized groups in society \cite{Somanathan2019May}. Automated counting in Philippines corresponded with dramatic reduction in result compilation time. \cite{philippines_transmission}.
%Countries including Argentina, Kenya, Bangladesh, Indonesia, Pakistan, Panama, Ghana, Kyrgyzstan and Kazakhstan have begun experimenting with election technology. Several countries, including Brazil, India, Namibia, Nigeria, and Venezuela, have adopted electronic voting machines, and Philippines and Mongolia have deployed automated counting systems. 

% negatives
However, there are frequent discrepancies and irregularities which undermine trust in technology. In 2017, the Supreme Court of Kenya nullified election results citing irregularities in the results transmission system \cite{Burke2017Sep}. In Pakistan, in 2018 the results transmission system broke down inexplicably on the night of elections, raising extreme suspicion \cite{Wasim2018Aug}. In Azerbaijan, introduction of a smartphone app in 2013 to report election results backfired when it released the election results the day before the actual election \cite{Bigg2013Oct}. In India, numerous incidents were reported in different polls where electronic voting machines `malfunctioned' by recording all votes in favour of the ruling party, no matter which choice the voter made \cite{india_mal}. In 2018, the introduction of untested voting machines in Democratic Republic of Congo was strongly opposed by opposition parties, and thousands of machines were subsequently destroyed in an act of arson \cite{Paravicini2018Dec}.

% how do we make sense of all this
Researchers have sought to explain these ``unintended consequences'' of election technology in terms of a ``fetishization of technology'' \cite{cheeseman2018digital}, or a silver bullet \cite{promiseandperils}, which distracts stakeholders from rigorous assessments and stringent checks and balances in the overall ecosystem. This lack of attention can render election processes even more vulnerable than before.

% role of e2ev voting systems
To situate the potential contribution of E2EVV systems, it is helpful to differentiate between electoral efficiency and transparency as two desirable yet distinct outcomes of using election technology \cite{yard2010direct}. Unfortunately, there is a marked tendency to prioritize efficiency over transparency, and favor a ``black box'' approach which concentrates trust ``away from the many'' and into the ''hands of the few''. We anticipate that E2EVV systems - by incorporating security and integrity as core design features of the system - can help redress this balance between electoral efficiency and public transparency. 

% calls in developing world
Similar sentiments have recently been voiced in the developing world, namely Brazil \cite{aranha2018good}, Pakistan \cite{ivtf_Report} \cite{haqelectronic}, and India \cite{cce_report}, where security professionals, researchers, and civil society organizations have urged election authorities to explore the adoption of E2EVV systems to restore credibility of electoral processes. Indeed, very recently in India, some 11 opposition parties unanimously passed resolutions affirming that the existing EVMs do not comply with ``democracy'' principles in that his or her vote is not verifiable \cite{11partiesres}.

%% file: 4_challenges.tex
\section{East is East and West is West: Misplaced Assumptions, Knowledge Gaps, and Other Challenges}

As we noted earlier, the design of most E2EVV systems is based on implicit assumptions which hold true for technologically advanced countries and do not necessarily translate easily to developing regions. A key goal in this section is to make these assumptions explicit by describing the challenges and knowledge gaps relevant to these environments. We divide these into four categories: structural constraints, social and political factors, human factors, and technical and operational concerns. We also include issues which are generic to adapting election technology and are well studied, but which may become more pronounced or take on added dimensions for the case of E2EVV systems.

\subsection{Structural Constraints}
\textbf{Shoestring Budgets:} Developing countries routinely suffer from severe financial constraints, and, due to large populations, election funding can take on disproportionate dimensions compared to other government priorities, such as poverty alleviation, and healthcare. For instance in 2018 general elections in Pakistan cost 21 billion PKR (~175 million USD) \cite{Kiani2018Jul}, comparable to the annual allocation for healthcare at 25 billion PKR (~208 million USD) and almost quarter the education spending at 97 billion PKR (~808 million USD) \cite{federalbudgetpak2018}.

Expensive technology interventions further strain these shoestring budgets. For instance, upgrading voting machines with paper trails in the Indian context cost 32 billion INR (~492 million USD)  \cite{3200crore}. Nationwide deployment of electronic voting machines in Pakistan are estimated to cost 350 billion PKR (2 billion USD), almost three quarters the national GDP \cite{Abbasi2021Dec}. These realities can foster undesirable trends: to quote the UN Secretary General, ``techniques and systems that might cause a State, in the conduct of its own elections, to be financially dependent on donors'' \cite{cheeseman2018digital}.

There is therefore a pronounced need to develop compact and minimalist E2EVV systems in a low-cost, sustainable manner, along the lines of India's famous voting machines. Perhaps, existing minimal FPGA-based E2EVV solutions like VoteBox Nano could be adapted for these settings \cite{oksuzoglu2009votebox}. A modular design approach would further maximize options to recycle components. 

The research community can contribute with open-source tools, software packages, libraries, kits, or hardware platforms to facilitate the development of such projects, similar to the ElectionGuard \cite{microsoft} effort or the wide availability of blockchain platforms like Hyperledger or Ethereum. 

Another promising direction is to develop E2EVV solutions which integrate with existing voting systems in developing countries. This is the design approach behind Scantegrity \cite{sherman2010scantegrity}. Recently, Mohanti et al. adapted risk limiting audits for Indian voting machines \cite{mohanty2019auditing}. \textbf{Is it possible to upgrade Indian or Brazilian machines in a similar cost-effective manner for verifiability?} 

\textbf{Resource and Infrastructure Woes:} Developing countries frequently suffer from resource shortages, including lack of essential equipment, IT systems, labs and storage facilities. Infrastructure problems include lack of utilities especially electricity, telecommunications, and Internet service. Expertise issues include poor access to IT expertise, quality technical support, and shortage of qualified polling staff. There is a dire need for indigenous capacity building and reforms for restructuring of broader management structures.

However there is encouraging evidence that technology can be creatively deployed within these constraints. India and Brazil's homegrown electronic voting machines are largely considered a success story. The under-banked in sub-Saharan Africa bypassed traditional banking and leap-frogged onto mobile money, accounting for 70\% of the global 1 trillion USD mobile money market \cite{Onyango2022May}. Alternatives need to be researched in response to specific challenges encountered in each country.

% bulletin board
To consider how infrastructure issues relate to E2EVV systems, we consider the specific example of the online bulletin board requirement. Whereas Internet access is ubiquitous in the developed world, in developing countries basic cell phone coverage and Internet access can be limited and unreliable due to network faults or traffic congestion. Our question becomes: \textbf{what kind of public bulletin board for vote verification could we offer in Asia and Africa where cell phone coverage is unreliable and an estimated 1.3 billion people still use dumb phones \cite{Cheney2018Mar}?} 

The popularity of text-messaging services (like SMS) may offer a way forward. These services have been successfully used in phone-based financial services, voter registration drives, social security programs, and mass vaccination efforts. Researchers have proposed SMS-based primitives including one time pads, return codes and transaction authentication numbers to harden remote voting systems \cite{backes2013using} which may potentially be leveraged for a bulletin board service over SMS. 

Infrastructure constraints may also be leveraged by malicious actors (e.g. a spoofing attack which misdirects voters to a fake bulletin board \cite{tjostheim2007case}). 
%[internet blackouts and bulletin boards?]

\subsection{Social and Political Factors}

\textbf{Electoral Fraud is Systemic:} It is well documented that developing countries often suffer from poor governance and endemic corruption \cite{olken2012corruption}, trends which also manifest in electoral practices. Vote buying, coercion, and suppression are commonplace: the 2013 Afrobarometer survey noted that 48 percent of voters in 33 African countries reported fearing violence during elections, whereas 16 percent reported being offered cash or goods for their vote. In Pakistan, in 2013, a watchdog body reported electoral irregularities at over 21,000 polling stations \cite{fafen2013}. In 2017, the Supreme Court of Kenya nullified election results citing irregularities in the results received over the results transmission system \cite{Burke2017Sep}. In 2017, in Venezuela, vendor Smartmatic disclosed that general results were ``manipulated'' and off by a count of at least 1 million \cite{Faiola2017Aug}. %Allegations of ``terrible fraud'' were a key justification for the recent military coup in Myanmar, where army officials claim to have identified some 10.5 million irregularities in the voter list used in the last general elections \cite{Goodman2021Feb}.

Persistent and systemic security threats in these environments necessitate additional security measures. \textbf{But what about attack scenarios caused by deploying E2EVV technology?} Poll workers can collect discarded receipts, voters may sell their receipts or surrender them on intimidation. Malicious parties could then manipulate the corresponding votes without fear of detection. 

A potential countermeasure is a verifiable encrypted paper audit trail (VEPAT) which incorporates additional checks performed by independent auditing authorities \cite{ryan2006verified}. These bodies could routinely verify the correspondence between the audit trail and receipts posted on the bulletin board. Solutions allowing voters to delegate the verification process to a trusted party \cite{simpson2018third} also merit investigation.

Another relevant concern: \textbf{how would E2EVV systems fare in environments where polling day security is lax and family voting, impersonation, and collusion are common?} These trends are well-documented in developing countries: a patriarch or another party obtains credentials of multiple legitimate voters and then casts votes on their behalf. Poll workers can cast votes on behalf of absentee voters. It is not surprising that the earliest election technology systems implemented in countries like Ghana, Nigeria, Kenya, DRC, Somaliland, Afghanistan are biometric voter verification systems. Following up on this, \textbf{can we integrate biometric checks with E2EVV systems in a binding way to provide enhanced security guarantees of voter identification, presence, and eligibility verifiability? Could these be done in a way that is universally verifiable?}

\noindent \textbf{The Politics of Perfection:} There is often lack of debate and rigorous analysis of election technology in developing countries. The Venezuelan government has described its electronic voting system as ``the most perfect voting system in the world''\cite{machin2016most}. The Indian Election Commission reacted angrily to reasonable security analysis of its voting machines \cite{indiaevm_book}. \textbf{Can developing countries who see their particular EVM systems as already ``perfect'' even begin to accept the need to evolve towards evidence-based elections and E2EVV systems? What kind of outreach effort would this entail?}

%existential threat

%The security guarantees of E2EVV systems become moot if the verification step is not undertaken, and the system is not auditable. If legislatures in developing countries cannot draft and pass laws that are sufficiently detailed to address both core and ancillary processes (such as risk limiting audits and electoral dispute resolution), E2EVV risks becoming nothing more than  the electoral equivalent of "security theatre" \cite{schneier_theater}. To quote Park et al ``\textit{Auditability} alone isn’t enough'', and ``must be accompanied by \textit{auditing} to be effective. \cite{park2021going}''. 

%For instance in India, group of political parties went to court to force ECI to audit paper trails, as paper trails have been merely ``ornamental''. Example of Brazil, audit very difficult, people lost interest [ask]. Introducing E2EVV thus requires an active and adequately resourced EMB that has a genuine will to implement transparency and effective audit procedures. 
 
\subsection{Human Factors}

%language diversity  aim for unity in diversity
\textbf{Linguistic and Cultural Diversity} Developing countries, marked by their linguistic diversity (dialects may change every few miles) ethnic diversity and varying cultural constructs \cite{weber2017societal}, require localization of both the voting system and accompanying receipts and verification mechanisms, which adds complexity to processes. People are often hesitant to carry out important transactions, especially ones involving  finances  in  an  unfamiliar  language. Language can potentially act as a barrier to election participation and disenfranchise certain voters \cite{pool1992multilingual}. \textbf{Can E2EVV systems cope with the sheer scope and scale of linguistic localization needed for many developing countries?}

%India alone has 22 national languages and to offer uniform vote casting and verification experience in each is necessary.
%Similarly, Urdu is the national language of Pakistan, however only 7\% of Pakistanis speak Urdu as their first language. Punjabi, Pashto, Sindhi, Saraiki, Balochi being the other prominent languages. Studies have observed that 

%Studies on ATM machines and mobile money systems in the developing world suggest that in multilingual information systems numerous translation inadequecies exist: This could be simply insufficient languages used for translation, linguistic issues such as total lack of translation, code switching, non-uniformity of language in the output of the system (i.e. the receipt), semantic, syntax and lexicon inadequecies.  

\noindent \textbf{What about Mental Models?} Mental models are essential to help foster public understanding of novel technology and customize interventions. Most voters think about a voting system first and foremost in terms of how to vote \cite{acemyan2015users}. Mental models for technology as well as attitudes and perceptions have been known to vary considerably in developing countries for certain applications \cite{hanel2018cross}. \textbf{What would mental models for E2EVV systems look like for voters in developing countries? How would these models vary, given the wide-ranging social and cultural diversity in these regions?}
%To advance acceptance, it is necessary to formulate mental models for E2EVV to help judges and EMB officials foster an understanding of this novel paradigm. \cite{zollinger2019user}. These can be targeted to help remove mismatch, impart voters clarity in their pre-existing mental models and overcome cognitive barriers.%\cite{kshetri2007barriers}.
% -usability  --verification - hash visualization?
%E2EVV systems have been known to confuse voters in developed countries \cite{acemyan2014usability}. %In E2EVV systems usability concerns are whether a voter can successfully cast their votes using these systems, and more importantly, whether or not they are able to undertake the verification process. 

\noindent \textbf{Usability - the highest hurdle?} Research points to correlation between low literacy (including digital literacy) and rejected ballots \cite{fujiwara}. Extending usability recommendations (for low-literacy voters in traditional electronic voting) to the E2EVV scenario is not trivial. A preliminary study involving Helios, Pret-a-Voter, and Scantegrity II found that it took almost twice as long to cast a vote, a significant number of voters failed to cast votes, and many did not realize their errors \cite{acemyan2014usability}. Voting success rates in developing countries will likely be lower. \textbf{How will this play out in developing countries with massive populations and already long queues, where voter or poll workers have been known to die of exhaustion \cite{indonesia_exhaustion}. Can E2EVV systems be designed to emphasize usability in low-literacy contexts?}?  

Moreover, recent testing and ‘live’ applications of E2E systems have resulted not just in consistently low rates of voter verification but even lower rates for those who actually report discrepancies \cite{moher2014diffusion}. Chipcase \cite{chipchase2005understanding} observe that non-literate populations avoid complex functions and this reinforces the assumption that if a step is optional, it will be skipped \cite{ellison2003upnp}.

\textbf{Can we develop technical solutions to simplify or automate the vote verification process in the context of developing countries?} Researchers have proposed solutions to make verifiability universal \cite{nandi2010stamp} \cite{ryan2011pret}, delegate it, or enable mass verification by bundling multiple receipts for batch verification \cite{bohli2009enhancing}. Could these be made more usable and practical for low-literacy users? Perhaps we could leverage research on textual key-fingerprint representations \cite{dechand2016empirical} and hash visualization \cite{azimpourkivi2020human} for this purpose. Research also shows that motivating messages can persuade voters to verify \cite{olembo2014voter}. \textbf{What sort of nudges or incentives could we devise to encourage voter verification in developing countries?}

\subsection{Governance and Operational Factors}

%-key management (problem even in developed countries)
\textbf{E-Governance and Digital Transformation:}

Developing countries often lack overarching institutional frameworks for governance, suffer from fragmentation and poor coordination of processes, and low uptake of digital technology. In the context of E2EVV systems, this can manifest in multiple ways. We consider the simple example of effective management of cryptographic credentials, recognized as problematic in trials of E2EVV systems. Logically, this problem will be more pronounced in developing countries.

Moreover, end-to-end verifiability and voter privacy are sensitive to human behaviour in the protocol. Errors in use of cryptography could result in exposure of critical data and undermine the integrity of the whole process. It would be helpful to characterize the set of behaviours under which security can be preserved and also higlight explicit scenarios where it fails \cite{kiayias2017ceremonies}. 
\textbf{How can we customize the key management and ceremonies to ensure separation of duties and principles of least privilege?} 

\textbf{Cybersecurity - Canaries in the Coalmine?}
Developing countries lag far behind in terms of capacities and resources for cybersecurity. For E2EVV, therefore,  vulnerabilities such as DDoS attacks on bulletin boards become more likely. Attacks on electoral information systems are on the increase and are often conducted by external or state actors with significant resources. The design of E2EVV systems must take this into account. \textbf{Can E2EVV systems be designed to be more "tamper-proof" (to use the infosec term) as well as "tamper-evident" (the elections term)?}

%Developing countries, starved of cybersecurity resources and expertise, where government and election websites are often vandalized, hacked, or spoofed to mislead citizens \cite{ECIhack} \cite{nadrafakewebsite}, and there is often a lack of cybersecurity teams, expertise, or protocols to detect and resolve these issues. In similar vein, there is a need to make the bulletin board resistant to DDoS attacks.

%EMBs need to develop a Cybersecurity strategy and undertake rigorous cyber hygiene exercises within all its departments. It needs to bring itself at par with the international standards with regards to cyber security. A step towards this is to obtain ISO-27001 certification. EMB’s are increasingly becoming the target for Cyber threats, with a 3-fold increase in attacks since 2015. One such incident involves a ransomware attack on the Caribbean EMB, which had to pay ransom in bitcoins to gain access to its data. A Computer Emergency and Response Team (CERT) should be established to handle any untoward incident.

%Similarly, the overall security environment needs to be assessed, including the possibility of state capture of all critical information systems under the mantle of "cybersecurity incident response")

\noindent \textbf{Legal Framework} In case of E2EVV systems, it should be legally binding on election management bodies (EMBs) to issue a receipt to every voter, upload all the receipts on a bulletin board within a stipulated time frame, verifying the results through the software provided to observers, making the software open source, sharing of public cryptographic parameters, conduct of Risk Limiting Audits. without which the security guarantees of E2EVV systems become moot. 
%The election day integrity measures, procedures also need to be updated in accordance with the new electronic system. 

%There needs to be a thorough analysis of what audit reports and forensic investigation data can be collected from the Electronic Voting System. In case the election results are challenged in court, what kind of evidence is admissible in court needs to be assessed apriori. 

As noted in the Brazil experience, ``important judicial decisions are not based on scientific research; they are often based on the personal opinions of judges who have no understanding of (election) technology.\cite{aranha2018good}'' Accordingly, for disputes involving technologies, defining the requirements for the admissibility of evidence, training the judiciary to handle the intricacies of E2EVV systems based digital evidence is of utmost importance. Moreover, comprehensive and high quality voter instruction is critical to uptake of a radically new system like E2EVV and typically falls under the auspices of the legal framework ensuring equal access. \cite{ali2016overview} Transparency and accountability are key to minimizing risk, and risk perception. 

Although it is not possible to have a generic,  one-size-fits-all set of requirements for E2EVV, we need to avoid the idea that all countries can do completely different things - fundamentally, technical requirements, and regulations, laws, and indeed constitutions, must deliver E2EVV systems that are fully compliant with universal principles.

\noindent \textbf{Toothless or Compromised EMBs and Ineffective Dispute Resolution} Many EMBs lack the regulatory teeth and political autonomy needed to ensure that incumbent government and political parties do not interfere in their duties, and they often operate under political influence and fear.\cite{undp_asia}. The lack of technical skills means there is unreliable implementation of technical protocols. 

%For instance, in India there have been numerous times when EVMs have ended up in private vehicles and homes. [citation needed]. In 2010, Philipines Compact Flash cards used for data storage (in which data can be transplanted), were used without applying seals, which introduced the possibility of physical tampering.

In developing countries, disputes over elections results often act as triggers for mass protests, violence, political deadlock and animosity, often times bordering civil war. This situation can be exacerbated when judicial mechanisms cannot resolve these disputes in a timely, fair and transparent manner \cite{norris2015contentious}. 

% corrupt EMBs
Another implicit assumption is that evidence of election malfeasance, if available from an E2EVV system, and provided to the authorities, would facilitate interventions by said authorities. There are various examples of EMBs in developing countries ignoring compelling evidence of electoral malfeasance. Moreover, in developing countries it is not uncommon for electoral processes to be politically controlled and for EMBs to be compromised. For instance, in Mozambique in 2014, the regime turned biometric registration into a technique of manipulation, suppressing registration in opposition areas by provisioning inadequate equipment and under-trained teams. In Kenya, over a million dead citizens were maintained in the voter register in an attempt to rig the polls \cite{kenyadeadvoters}. Venezuelan election results were internally `manipulated' by at least 1 million votes \cite{Faiola2017Aug}.

In such situations, E2EVV systems, with their rigorous security guarantees, may well be perceived as an existential threat. \textbf{In this regard, to what degree could E2EVV systems be corrupted in a compromised ecosystem? Moreover, could E2EVV systems possibly be developed to offer guarantees against a compromised ecosystem? Could solutions be developed to render these ecosystem issues transparent as well?}

\noindent \textbf{Electoral Integrity Theatre} The security guarantees of E2EVV systems become moot if the verification step is not undertaken, and the system is not auditable. If legislatures in developing countries cannot pass and enforce laws that are sufficiently detailed to address both core and ancillary processes, E2EVV risks becoming nothing more than  the electoral equivalent of "security theatre" \cite{schneier_theater}. To quote Park et al ``\textit{Auditability} alone isn’t enough'', and ``must be accompanied by \textit{auditing} to be effective. \cite{park2021going}''. \textbf{There is therefore a need for public awareness on this issue and devising satisfactory mechanisms, policies, and legislation to enforce electoral integrity checks.}

\noindent \textbf{Belt and Braces} E2EVV systems need to be made resilient with backup 'belt and braces' mechanism. For instance, Star-Vote is a system which incorporates Risk Limiting Audits to an E2EVV system. Risk Limiting Audits have even been devised to cater to on-ground realities in India. It is essential to work in close engagement with existing systems on the ground.

%% file: 5_wayforward.tex
\section{Conclusion} 
%Given the poor uptake of E2EVV mechanisms in developed countries, and in light of the challenges and explicit assumptions above, significant further research is required to achieve the increase necessary to make E2EVV a valid option. 
%We offer as suggestions for further research: (1) Expanding the concepts in \cite{backes2013using} to use SMS and USSD rather than mobile internet, thereby opening their use to "dumb" phones. (2) Explore ways to make coercion evidence \cite{KulkyNeumannHumanFactors} impossible to conceal by an untrusted or nontransparent EMB. (3) Explore the impact on E2E-VV models when the concept of ceremonies \cite{kiayias2017ceremonies} and greater user input \cite{KulkyNeumannHumanFactors} are included. (4) Can E2E-VV survive a Mexican Standoff? (Where each stakeholder regards all other stakeholders as compromised?)
%Ronans alternative, shorter last paragraph
%Suggestions for further research: (1) Examine voter-friendly alternatives (like SMS and USSD) for E2E-VV. (2) Research "transparent by default" E2E-VV to force EMBs to act on coercion evidence\cite{KulkyNeumannHumanFactors}, (3) Design an E2E-VV system in a developing country but do so with genuine voter input.
%Here's an alternative Conclusion, as requested.... Feel free to delete with great prejudice...
Despite formidable recent efforts to portray elections management in the United States as dysfunctional and corrupt, the reality in developed countries is of well-resourced EMBs, reliable infrastructure, competent staff, reliable dispute resolution mechanisms, digitally literate voters and empowered civil society and media stakeholders. In contrast, we have outlined systemic problems in most developing countries in most of those aspects. Accordingly, the underlying assumptions of most E2EVV systems mean that implementation of E2EVV in developing countries is an uphill task. Some of the solutions we have discussed thus far even clash with one another. For example, if biometrics need to be introduced to deal with corruption and fraud, that would increase the cost, thereby counteracting efforts to reduce that cost. Significant research with an explicit focus on developing country contexts is needed in order to bridge this gap.  Given the potential benefits of E2EVV, we believe this pivot is well justified. We hope our paper is a catalyst in this regard.